\documentclass[fleqn,twoside]{article}
\usepackage{espcrc2}
\usepackage{epsfig}
\usepackage{latexsym}
\usepackage{amssymb}
\usepackage{amsfonts}

\readRCS
$Id: espcrc2.tex,v 1.2 2004/02/24 11:22:11 spepping Exp $
\usepackage{graphicx}
\usepackage[figuresright]{rotating}

\newcommand{\AmS}{{\protect\the\textfont2
  A\kern-.1667em\lower.5ex\hbox{M}\kern-.125emS}}

\title{
$S$-bases as a tool to solve reduction problems
for Feynman integrals}

\author{A.V.~Smirnov\address{Scientific Research Computing Center of
Moscow State University, Moscow 119992, Russia}\thanks{Supported by
the Russian Foundation for Basic Research
through grant 05-01-00988.}
and V.A.~Smirnov\address{Nuclear Physics Institute of
Moscow State University,
Moscow 119992, Russia}\thanks{Talk given at the International
Workshop `Loops and Legs in
Quantum Field Theory' (April 23--28, 2006, Eisenach, Germany).
Supported by the Russian Foundation for Basic Research through grant
05-02-17645.}}

\newcommand{\be}{\begin{equation}}
\newcommand{\ee}{\end{equation}}
\newcommand{\bea}{\begin{eqnarray}}
\newcommand{\eea}{\end{eqnarray}}
\newcommand{\al}{\alpha}

\newcommand{\dl}{\delta}

\newcommand{\ep}{\varepsilon}
\newcommand{\sg}{\sigma}

\newcommand{\pa}{\partial}
\newcommand{\dd}{\mbox{d}}

\newcommand{\nn}{\nonumber}

\newcommand{\SA}{\left\{ \begin{array}{ll}}
\newcommand{\Sa}{\left[ \begin{array}{ll}}
\newcommand{\FA}{\end{array}\right.}

\newcommand{\leftdot}{$\!\!\!\mbox{\bf{.}}\,\,$}
\newtheorem{tr}{Theorem}
\newcommand{\BTh}{\begin{tr}\leftdot}
\newcommand{\ETh}{\end{tr}}
\newtheorem{lmm}{Lemma}
\newcommand{\BLm}{\begin{lmm}\leftdot}
\newcommand{\ELm}{\end{lmm}}
\newtheorem{deff}{Definition}
\newcommand{\BDf}{\begin{deff}\leftdot}
\newcommand{\EDf}{\end{deff}}
\newtheorem{stttt}{Step}
\newcommand{\BStt}{\begin{stttt}\leftdot}
\newcommand{\EStt}{\end{stttt}}
\newenvironment{proof}{$\blacktriangleleft$ }{$\blacktriangleright$
\indent}
\newcommand{\BPr}{\begin{proof}}
\newcommand{\EPr}{\end{proof}}
\newcommand{\BZM}{\begin{zamet}}
\newcommand{\EZM}{\end{zamet}}

\begin{document}

\begin{abstract}
We suggest a mathematical definition of the notion of master
integrals and present a brief review of algorithmic methods to solve
reduction problems for Feynman integrals based on integration by
parts relations. In particular, we discuss a recently suggested
reduction algorithm which uses Gr\"obner bases. New results
obtained with its help for a family of three-loop Feynman
integrals are outlined.
\vspace{1pc}
\end{abstract}

\maketitle

\section{Introduction}

In the framework of perturbation theory quan\-tum-theoretical
amplitudes are written in terms of Feynman integrals that
are constructed according to Feynman rules.
After a tensor reduction based on projectors is performed,
each Feynman graph generates various scalar Feynman integrals
with the same structure of the integrand and with various
powers of propagators ({\em indices}):
\bea
  F(a_1,\ldots,a_n) &=&
  \int \cdots \int \frac{\dd^d k_1\ldots \dd^d k_h}
  {E_1^{a_1}\ldots E_n^{a_n}}\,.
  \label{eqbn}
\eea
Here $k_i$, $i=1,\ldots,h$, are loop momenta
and the denominators $E_r$ are either quadratic or linear with respect
to the loop momenta $p_i=k_i, \; i=1,\ldots,h$ or
to the independent external momenta $p_{h+1}=q_1,\ldots,p_{h+N}=q_N$ of the graph.
Irreducible polynomials in the numerator can be represented as
denominators raised to negative powers.
Usual prescriptions $k^2=k^2+i 0$, etc. are implied.
The dimensional regularization
with $d=4-2\ep$ is assumed.

In today's perturbative calculations, when one needs
to evaluate thousands or millions of Feynman integrals (\ref{eqbn}), a
well-known traditional strategy is
to apply integration by parts (IBP) relations
\cite{IBP} and some other relations
in order to find an
algorithm that expresses any given Feynman integral as a linear
combination of some {\it master} integrals.
Then the whole problem of evaluation of Feynman integrals of a
given family is decomposed into two parts: constructing a reduction procedure
and evaluating master integrals.


It turns out, however, that there is no common definition of
master integrals. Practically, after solving the reduction
problem for a given family,
{\em we know that these are master integrals because we see them}.
There are examples where different authors consider different master
integrals for the same class of Feynman integrals.
In the next section, we are going to suggest 
a definition of this notion. 
In Sections~3 and~4, we give a short review of algorithmic methods
to solve recursion problems in the framework of this definition.
In particular, we characterize in Section~4 our algorithmic method
\cite{2S,AS06} which uses so-called $s$-bases and outline new
results obtained with its help for a family of three-loop Feynman
integrals.

\section{The notion of master integrals}
%
%


To define the notion of master integrals 
we need some definitions.
Feynman integrals (\ref{eqbn}) can be considered as elements of
the field of functions $\mathcal F$ of $n$ integer
arguments $a_1,a_2,\ldots,a_n$. 
This is an infinitely dimensional linear
space. The simplest basis
of this space is the set of elements $H_{a_1,\ldots,a_n}$, where
$H_{a_1,\ldots,a_n}(a'_1,\ldots,a'_n)=
\dl_{a_1,a'_1}\ldots\dl_{a_n,a'_n}\;.$

One usually considers
some set of relations between Feynman
integrals. Formally, they are elements of the adjoint vector
space $\mathcal F^*$, i.e. the linear functionals on $\mathcal F$,
so that, for any $r\in {\mathcal F^*}$, there is a corresponding value
$\langle r, f \rangle$ for any given $f\in {\mathcal F}$ .
The simplest basis
of this space is the set of elements $H^*_{a_1,\ldots,a_n}$
which are defined as follows:
\bea
\langle H^*_{a_1,\ldots,a_n}, f \rangle
= f(a_1,\ldots,a_n) \;.
\nn
\eea

There are several types of commonly used relations.
The first and the main type consists of the
integration by parts relations \cite{IBP}
\bea
\int\ldots\int \prod_{i'}\dd^d k_{i'}
\frac{\pa}{\pa k_i}\left( p_j
\prod_{l=1}^n E_l^{-a_l}
\right)   =0  \;.
\label{RR}
\eea
After differentiating, the scalar products
$k_i\cdot k_j$ and $k_i\cdot q_j$
are expressed linearly in terms of the factors $E_i$ of the denominator,
and one obtains the IBP relations in the following form:
\begin{equation}
\sum \al_i F(a_1+b_{i,1},\ldots,a_n+b_{i,n})
=0\,,
\label{IBP}
\end{equation}
Now one can substitute all possible $(a_1,a_2,\ldots,a_n)$ on the left-hand sides
of (\ref{IBP}) and obtain an infinite set of elements of $\mathcal F^*$.

In addition to the IBP relations one often considers the so-called
Lorentz-invariance identities \cite{GR1}.
It is not yet clear whether they follow from IBP
relations but they turn out to be useful.

Another type of relations comes from the symmetries of the
given family of integrals. Typically they have the following form:
\bea
F(a_1,\ldots,a_n)=(-1)^{\sum d_i a_i}
F(a_{\pi(1)},\ldots,a_{\pi(n)})\,, \nn
\eea
where $d_i$ are fixed and are equal to either one or zero, and $\pi$ is a
permutation. Again, after substituting all possible $a_i$
we obtain an infinite set of elements of $\mathcal F^*$.

Then one takes into account boundary conditions,
i.e. the conditions of the following form:
\be
F(a_1,\ldots,a_n)=0\mbox{ when }a_{i_1}\leq 0,\ldots a_{i_k}\leq 0
\label{boundary}
\ee
for some subset of indices.

One more type of relations corresponds to parity conditions.
For example, Feynman integrals can
be zero if the sum of some subset of indices is odd and each index
is nonpositive.

After having fixed a set of relations
we can generate by them an
infinitely dimensional vector subspace ${\mathcal R} \subset {\mathcal F^*}$.
Now one considers the set of solutions of all those relations, that is
the intersection of the kernels of all functionals
$r \in {\mathcal R}$. This is a vector subspace of $\mathcal F$,
that will be denoted with $\mathcal S$. A Feynman integral
considered
as a function of the integer variables $a_1,\ldots,a_n$ is an element of the
space $\mathcal S$ for it satisfies the IBP relations and other
relations mentioned above.
Formally,
${\mathcal S} =\{ f\in {\mathcal F}: \langle r,f\rangle =0
\; \forall\; r\in {\mathcal R}\}$.

The dimension of $\mathcal S$ might be infinite. However, for all
known classes of Feynman integrals, where the reduction problem
has been solved, it appears to be finite. It looks like it can be
a non-trivial mathematical problem to prove that this holds for all
families of Feynman integrals.

When talking about expressing one Feynman integral by
another, it is usually assumed that we consider the consequences of
relations $\mathcal R$. Let us say that an integral
$F(a_1,\ldots,a_n)$ can be expressed via some other integrals
$F(a^1_1,\ldots,a^1_n)$, $\ldots$, $F(a^k_1,\ldots,a^k_n)$ if there
exists 
$r\in {\mathcal R}$ such that
\bea
\langle r,F\rangle
=F(a_1,\ldots,a_n)
\nn \\ && \hspace*{-25mm}
+\sum k_{a'_1,\ldots,a'_n}F(a'_1,\ldots,a'_n)
\;.
\label{expressed}
\eea


Let us turn to the notion of irreducibility of Feynman integrals. Suppose
we have two integrals $F(a_1,\ldots,a_n)$ and
$F(a'_1,\ldots,a'_n)$ that can be expressed one by another, for
example, due to a symmetry of the diagram.
Of course, it is reasonable to choose only one of them as a master
integral.
However there seems to be nothing natural in this
choice, for they are equivalent. So, even having fixed a set of
relations, we do not have enough information to define master
integrals. The only thing we know that their number is equal
to the dimension of $\mathcal S$.

Therefore, to define master (or, irreducible) integrals, we
need to choose a certain priority between the points
$(a_1,\ldots,a_n)$, formally, to
introduce a complete ordering on them (that will be denoted with
the symbol $\prec$ and named as \textit{lower}).
There are different ways to do that but at least
it looks natural to have
simpler integrals corresponding to the minimal elements in this
ordering.
We shall introduce an ordering in two steps.
First of all, let us realize that the Feynman integrals are
simpler, from the analytic point of view, if they have more
non-positive indices. In fact, in numerous examples of solving
of IBP relations by hand, the natural goal was to reduce indices to
zero or negative values.\footnote{Moreover, the reduction procedure often
stopped once one arrived at integrals which are already sufficiently
simple, in particular, when they can be expressed analytically in
terms of gamma functions for general $d$.}
The big experience reflected in many papers
has led to the natural idea to decompose the whole region of the
integer indices which we call {\em sectors}.\footnote{
This decomposition is also standard in the
so-called Laporta's algorithm
\cite{Lap} which is briefly characterized below.
}

Let us consider the set $\cal D$ of elements $\{d_1,\ldots,d_n\}$
called \textit{directions},
where all $d_i$ are equal to $1$ or $-1$. For any given direction
$\nu=\{d_1,\ldots,d_n\}$, we consider the region
$\sg_\nu = \{ (a_1,\ldots,a_n): (a_i - 1/2)  d_i > 0\}$
and call it {\em a sector}.
In other words, in a given sector, the indices corresponding to $\pm 1$
are positive (non-positive).
Obviously the union of all sectors contains all integer points
in the $n$-dimensional vector space and the intersection of any two
sectors is an empty set. We will say that a direction $\{d_1,\ldots d_n\}$ is lower than
$\{d'_1,\ldots d'_n\}$ if $d_1\leq d'_1,\ldots,d_n\leq d'_n$
and they are not equal. The
same is said about the corresponding sectors.

It looks natural to define that $F(a_1,\ldots,a_n)\succ F(a'_1,\ldots,a'_n)$
if the sector of $(a'_1,\ldots,a'_n)$ is lower than the sector of
$(a_1,\ldots,a_n)$.
Moreover it is also natural to assume that the `corner point'
$((d_1+1)/2,\ldots,(d_n+1)/2)$ (those numbers are either ones or zeros)
of the sector $\sigma_{\{d_1,\ldots,d_n\}}$ is lower than
all other points of this sector.

To define an ordering completely
we introduce it in some way inside the sectors.
After this, we can define what a master integral is. It is such an
integral $F(a_1,\ldots,a_n)$ that there is no
element $r\in {\mathcal R}$
acting on $F$ according to relation (\ref{expressed})
such that all the points
$(a'_1,\ldots,a'_n)$ are lower than $(a_1,\ldots,a_n)$.


As it has been noted in \cite{Bai}, one can prove that a given
integral $F(a_1,\ldots,a_n)$ is a master integral 
by constructing a function $c\in\mathcal F$ such that (*)
\\
i) it satisfies all relations $\mathcal R$ and therefore
$c\in\mathcal S$;
\\
ii) $c(a_1,\ldots,a_n)=1$;
\\
iii) $c(a'_1,\ldots,a'_n)=0$ for all lower $(a'_1,\ldots,a'_n)$.
\\
To do this 
one can use
the basic parametric representation (\ref{Basic}) of the Baikov's
method \cite{Bai} which is discussed below.


\section{Algorithmic methods of solving \\ reduction problems}

Now we can give a short overview of the known algorithmic methods from the
standpoint of our definitions.

The Laporta's algorithm \cite{Lap}
works with systems of equations for individual Feynman integrals.
Let $\mathcal F_t$ be the set the subspace of $\mathcal F$
generated by $H_{a_1,\ldots,a_n}$ where $\mid a_i \mid \leq t$ and
$\mathcal R_t$ be the intersection of $\mathcal R$
with the subset of $\mathcal F^*$ generated by
$H^*_{a_1,\ldots,a_n}$ where $\mid a_i \mid \leq t$.
The limit of the difference between the dimensions of $\mathcal F_t$
and $\mathcal R_t$ when $t$ tends to infinity is the dimension
of 
$\mathcal S$, so that there is a certain $t$ such that
$\mathcal R_t$ has `enough' relations to express any given
integral $F(a_1,\ldots,a_n)$ with $\mid a_i \mid \leq t$ as
a linear combination of master integrals with the use of those
relations. So, the method is based on finding such a $t$ and
solving a huge system of linear equations.

There is a public version \cite{AnLa}
of the implementation of this algorithm as well as a number
of private versions.

The basic tool of the Baikov's method \cite{Bai,ST}
(see also Chapter~6 of \cite{EFI})
is the 
representation
\begin{eqnarray}
  \int\ldots\int \frac{\dd x_1 \ldots \dd x_n}{x_1^{a_1} \ldots
    x_n^{a_n}} \left[P ( x_1, \ldots,x_n)\right]^{(d-h-1)/2} \,,
\label{Basic}
\end{eqnarray}
where $P$ is constructed for a given family of integrals according
to some rules.
This representation is used to prove that an integral with
$a_{i1},\ldots,a_{in}=0$ is a master integral. Indeed, such a
function automatically satisfies two of the three points of (*)
and it is left to check whether it does not vanish at
$(a_{1},\ldots,a_{n})$.



Suppose that we know that the integrals $F(A^j)$ with 
$A^1=(a^1_1,\ldots,a^1_n)$, $\ldots$, $A^k= (a^k_1,\ldots,a^k_n)$
are master integrals for we have constructed the corresponding
solutions of this type $C_{(a_1,\ldots,a_n)}$ satisfying the
conditions (*).
%
%
%
These functions form a subbasis of the solution space $\mathcal S$.
Let us suppose that $F$ can be represented as
\begin{equation}
F=\sum_i k_i C_i \;.
\label{linear}
\nn
\end{equation}
(This is always true if we know the complete set of master
integrals.) Then one substitutes all $A_j$  and solves an upper-triangle system
of 
equations
\[
F(A_j)=\sum_i k_i C_i(A_j)\mbox{ for }i\leq j\,.
\]
%
so the coefficients $k_i$ are expressed in
terms of the values $F(A_i)$.
Then the knowledge of both $k_i$ and $C_i$ provides the possibility
to calculate any given Feynman integral of this class.
In fact it can happen that there is
more than one master integral in a given sector, therefore one
should consider also points different from sector corners.

This method has not yet been analyzed from the mathematical point of
view. After constructing a set of master integrals it is not clear
whether this is a complete set. Still one can always perform various
crucial checks and then understand that the recursion problem was
solved indeed.

\section{Using Gr\"obner bases and $s$-bases 
}

Tarasov suggested the idea \cite{Tar2} to apply
Gr\"obner bases \cite{Buch} to solve
reduction problems for Feynman integrals\footnote{As an application of
this approach, the solution of
the reduction problem for two-loop self-energy diagrams
with five general masses was obtained in \cite{Groe}, with an
agreement with an earlier solution  \cite{Tar98}.
The solution of
the reduction problem for massless two-loop off-shell vertex
diagrams (which was first obtained in \cite{BGM} within
Laporta's algorithm) was reproduced in \cite{JeTa}.}.
In his approach, IBP relations are reduced to differential
equations.
An attempt to use Gr\"obner bases associated with the shift operators
was made in \cite{Gerdt}.

In the recently introduced method \cite{2S,AS06}
which uses so-called $s$-bases (Gr\"obner-type bases),
the ordering in a given sector
should be chosen more strictly.
Let us fix a sector $\sigma_{(c_1,\ldots,c_n)}$ and
take $p_i=(c_1+1)/2$. Then let us say that a (multi)-degree of a point
$(a_1,\ldots,a_n)\in\sigma_{(c_1,\ldots,c_n)}$ is
$(c_1(a_1-p_1),\ldots,c_n(a_n-p_n))$. The degree of the corner of
the sector $(p_1,\ldots,p_n)$ is obviously equal to
$(0,\ldots,0$).
The degrees form the semi-group $\mathbb N^n$
(with respect to
$(b_1,\ldots,b_n)+(b'_1,\ldots,b'_n)=(b_1+b'_1,\ldots,b_n+b'_n)$).
An ordering on the points is naturally extended to the ordering of
degrees. We will say that an ordering on $\mathbb N^n$
(denoted with the symbol $\succ$) is
\textit{proper} if
\\
i) for any $a\in \mathbb N^n$ not equal to $(0,\ldots 0)$
one has $a\succ (0,\ldots 0)$
\\
ii) for any $a,b,c\in \mathbb N^n$ one has $a\prec b$
if and only if $a+c\prec b+c$.

Now we are allowed to take any ordering such that the
corresponding ordering on degrees is proper for each sector.
This corresponds to the basic idea of the method to introduce the
algebra acting on $\mathcal F$ and to work
with operators and their degrees. Let us outline it briefly.
This algebra $\mathcal A$ is
generated by shift operators $Y_i$ and multiplication operators
$A_i$ that act via
\bea
(Y_i\cdot F)(a_1,\ldots,a_n)=F(a_1,\ldots,
a_i+1,
\ldots,a_n)\,,
\nn \\ \nn
(A_i \cdot F)(a_1,\ldots,a_n)=a_i F(a_1,\ldots,a_n)\,. \nn
\eea
The IBP relations are interpreted as elements of this algebra
and generate an ideal $\mathcal I$.
There is an algorithm \cite{2S,AS06}
designed to build a set of special bases
of this ideal, allowing to reduce any integral to master
integrals. To elucidate, any element of this ideal corresponds
to an infinite number of relations $r\in\mathcal R$ obtained by
substituting different $a_i$.
Obviously, 
\bea
F(a_1,\ldots,a_n)
&& \nn \\ && \hspace*{-15mm}
=\left(\prod_{j=1}^n
(Y^{c_j}_j)^{c_j(a_j-p_j)}\cdot F\right)(p_1,\ldots,p_n)\,,
\nn
\eea
and the exponents at $Y^{c_j}_j$ form exactly the degree of
$(a_1,\ldots,a_n)$. Hence expressing an integral with lower
integrals corresponds to the idea of the method to build the
elements of $\mathcal I$ with lowest possible degrees.

If one has constructed the $s$-bases for all sectors, the reduction
algorithm can be applied. It expresses any integral as a linear
combination of a certain number of integrals, that could not be
reduced by the method.


This algorithm can work for problems
with many indices.
In \cite{GSS}, it was applied to three-loop
Feynman integrals 
\bea
F(a_1,\ldots,a_9)
&=&\int \int\frac{\dd^d k\, \dd^d l  \dd^d r}{
(-2 v\cdot k)^{a_1} (-2 v\cdot l)^{a_2}}
\nn \\ && \hspace*{-25mm}
\times
\frac{1}{(-k^2)^{a_3} (-l^2)^{a_4} [-(k-l)^2]^{a_5}(-r^2+m^2)^{a_6}}
\nn \\ && \hspace*{-25mm}
\times\int  \frac{ (2 v\cdot r)^{-a_9}}{
[-(k+r)^2+m^2]^{a_7} [-(l+r)^2+m^2]^{a_8}}\;.
\nn
\eea
Here $a_9\leq 0$, $v$ is the HQET quark velocity. 
The integrals are symmetric with respect to
$(1\leftrightarrow2,3\leftrightarrow4,7\leftrightarrow8)$
and vanish if one of the following sets is with non-positive indices:
$a_{5,7},a_{5,8},a_{6,7},a_{6,8},a_{7,8}, a_{3,4,6}$.
The algorithm of \cite{2S,AS06} has revealed the following
master integrals:
\bea
I_1&=&F(1,1,0,1,1,1,1,0,0)\;,\nn\\
I_2&=&F(1,1,1,1,0,0,1,1,0)\;,\nn\\
I_3&=&F(1,1,0,0,0,1,1,1,0)\;,\nn\\
I_4&=&F(0,1,1,0,1,1,0,1,0)\;,\nn\\
\bar{I}_4&=&F(-1,1,1,0,1,1,0,1,0)\;,\nn\\
I_5&=&F(0,0,0,1,1,1,1,0,0)\;,\nn\\
I_6&=&F(0,1,0,0,0,1,1,1,0)\;,\nn\\
I_7&=&F(0,1,0,0,1,1,1,0,0)\;,\nn\\
\bar{I}_7&=&F(0,2,0,0,1,1,1,0,0)\;,\nn\\
I_8&=&F(0,0,0,0,0,1,1,1,0)\;.\nn
\eea

These are some examples of reduction:
\bea
F(1, \ldots,1, 0)=
\frac{3d-10}{d-5}\left\{
-\frac{3(d-4) }{8 (2d-9)} I_1 \right.
&& \nn \\ && \hspace*{-55mm}
- \frac{3(d-4) }{16 (2d-9)} I_2
- \frac{(d-3) (3d-8 )}{8 (3d-13)(3d-11)} I_3
\nn \\ && \hspace*{-55mm}
- \frac{3(d-2)(3d-11) (3d-8 )}
{64 ( 2d-9)( 2d-7)( 3d-13)} \bar{I}_4
\nn \\ && \hspace*{-55mm}
+ \frac{9( d-4)( d-2) (3d-8)}
{64 (2d-9)(2d-7 )( 3d-13)} I_5
\nn \\ && \hspace*{-55mm} \left.
- \frac{3 ( 3d-8)}{32 ( 2d-9)( 2d-7)} \bar{I}_7
\right\}
\;,
\nn \\ && \hspace*{-69mm}
F(1, \ldots,1, -1)=
\frac{3(  d-3)(  3d-11)}{16(  d-5)(  d-4)(  2d-9)}I_4
\nn \\ && \hspace*{-69mm}
-\frac{(  d-2)(  2d-7)(  2d-5)}{8(  d-3)(  2d-9)(  3d-13)}I_6
\nn \\ && \hspace*{-69mm}
 - \frac{3( 2d-7)^2(  2d-5)(  3d-11)(  3d-7)}
{256(d-4)^2(  d-3)(  2d-9)} I_7 \;.
\nn
\eea

Preliminary checks have shown that the present algorithm
can work for large $n$,
e.g., for $n=12$.
There are various interesting practical and mathematical
problems connected with method.
We hope to solve them in the future and improve our algorithm for more sophisticated
calculations.


\begin{thebibliography}{9}
\bibitem{IBP}
K.G.~Chetyrkin and F.V.~Tkachov,
Nucl. Phys. B 192 (1981) 159.

\bibitem{2S}
A.V.~Smirnov and V.A.~Smirnov, JHEP 01 (2005) 001;
V.A.~Smirnov, hep-ph/0601268.

\bibitem{AS06}
A.V.~Smirnov, JHEP 04 (2006) 026.
See also
http://www.srcc.msu.ru/nivc/about/lab/

\noindent lab4\_2/index\_eng.htm.

\bibitem{GR1}
T.~Gehrmann and E.~Remiddi, Nucl. Phys. B 580 (2000) 485.

\bibitem{Lap}
S.~Laporta and E.~Remiddi, Phys. Lett. B 379 (1996) 283;
S.~Laporta, Int. J. Mod. Phys. A 15 (2000) 5087;
T.~Gehrmann and E.~Remiddi,
Nucl. Phys. B 601 (2001) 248;
Nucl. Phys. B 601 (2001) 287.

\bibitem{AnLa}
C.~Anastasiou and A.~Lazopoulos, JHEP 0407 (2004) 046.

\bibitem{Bai}
P.A.~Baikov, Phys. Lett.  B 385  (1996) 404;
Nucl.~Instrum.~Methods A 389 (1997) 347;
Phys. Lett. B 474 (2000) 385;
Phys. Lett. B 634 (2006) 325.

\bibitem{ST}
V.A.~Smirnov and M.~Steinhauser,
Nucl. Phys.  B 672 (2003) 199.

\bibitem{EFI}
V.A.~Smirnov, {\it Evaluating Feynman integrals},
STMP 211; Springer, Berlin, Heidelberg, 2004).

\bibitem{Buch}
B.~Buchberger and F.~Winkler (eds.) {\it
Gr\"obner Bases and Applications}, (Cambridge University Press, 1998).

\bibitem{Tar2}
O.V.~Tarasov, Acta Phys. Polon. B  29 (1998) 2655.

\bibitem{Groe}
O.V.~Tarasov,
Nucl. Instrum. Meth. A  534 (2004) 293.

\bibitem{Tar98}
O.V.~Tarasov, Nucl. Phys. B 502 (1997) 455.

\bibitem{BGM}
T.G.~Birthwright, E.W.N.~Glover and P.~Marquard,
JHEP 0409 (2004) 042.

\bibitem{JeTa}
F.~Jegerlehner and O.V.~Tarasov, hep-ph/0510308;
hep-th/0602137.

\bibitem{Gerdt}
V.P.~Gerdt, Nucl. Phys. B (Proc. Suppl), 135 (2004) 2320;
math-ph/0509050;
V.P.~Gerdt and D.~Robertz, cs.SC/0509070.

\bibitem{GSS}
A.G.~Grozin, A.V.~Smirnov and V.A.~Smir\-nov, to be published.
\end{thebibliography}
\end{document}